\documentclass[
    aps,
    prd,
    superscriptaddress,
    twocolumn,
    a4paper,
    floatfix,
    longbibliography
]{revtex4-2}

\usepackage[american]{babel}
\usepackage[utf8x]{inputenc}

\usepackage{grffile} 

\usepackage{newtxtext,newtxmath}
\usepackage{microtype}

\usepackage{amsmath}
\usepackage{amssymb}
\usepackage{bbm}
\usepackage{mathtools}
\usepackage{dsfont}
\usepackage{braket}
\usepackage{cancel}
\usepackage{slashed}

\usepackage{graphicx}
\usepackage[svgnames]{xcolor}

\usepackage{siunitx} 

\usepackage{ragged2e}
\usepackage{array}
\usepackage{tabularx}
\usepackage{booktabs}

\definecolor{cset-aps-blueberry}{RGB}{28,128,158}
\definecolor{cset-aps-blue}{RGB}{46,44,184}
\definecolor{cset-aps-turquoise}{RGB}{0,67,88}
\definecolor{cset-aps-limegreen}{RGB}{190,219,67}
\definecolor{cset-aps-green}{RGB}{31,138,112}
\definecolor{cset-aps-yellow}{RGB}{255,225,25}
\definecolor{cset-aps-orange}{RGB}{253,116,0}
\definecolor{cset-aps-red}{RGB}{219,0,43}

\usepackage{tikz}

\usepackage{pgfplots}
\pgfplotsset{%
    every axis legend/.append style={%
        cells={anchor=west},
        at={(0.96,0.04)},
        anchor=south east,
        font=\scriptsize,
        },
    every axis/.append style={%
        yticklabel style={%
            /pgf/number format/fixed zerofill,
            /pgf/number format/precision=2},
        },
    width= \textwidth,
    height=8cm,
    xmajorgrids=true,
    xminorgrids=false,
    minor x tick num=1,
}
\usepgfplotslibrary{external}

\usetikzlibrary{decorations}
\pgfplotsset{compat=1.17}

\usepackage{pict2e,picture}

\makeatletter
\DeclareRobustCommand{\Arrow}[1][]{%
\check@mathfonts
\if\relax\detokenize{#1}\relax
\settowidth{\dimen@}{$\m@th\rightarrow$}%
\else
\setlength{\dimen@}{#1}%
\fi
\sbox\z@{\usefont{U}{lasy}{m}{n}\symbol{41}}%
\begin{picture}(\dimen@,\ht\z@)
\roundcap
\put(\dimexpr\dimen@-.7\wd\z@,0){\usebox\z@}
\put(0,\fontdimen22\textfont2){\line(1,0){\dimen@}}
\end{picture}%
}
\makeatother

\usepackage{hyperref}
\hypersetup{%
    colorlinks=true,
    linkcolor={cset-aps-red},
    linkbordercolor={cset-aps-red},
    filecolor={cset-aps-orange},
    filebordercolor={cset-aps-orange},
    citecolor={cset-aps-blue},
    citebordercolor={cset-aps-blue},
    urlcolor={cset-aps-green},
    urlbordercolor={cset-aps-green},
    menucolor={cset-aps-limegreen},
    menubordercolor={cset-aps-limegreen},
    breaklinks=true,
    pdfborderstyle={/S/U/W 2},
    pdfpagemode=UseOutlines,
    pdfstartpage={1},
}

\newcommand{\ee}{\text{e}}
\newcommand{\ii}{\text{i}}

\usepackage{lipsum}

\newcommand{\affHAN}{\address{Institut f{\"u}r Quantenoptik, Leibniz Universit{\"a}t Hannover, Welfengarten 1, D-30167 Hannover, Germany}}
\newcommand{\affULM}{\address{Institut f{\"u}r Quantenphysik and Center for Integrated Quantum Science and Technology (IQ\textsuperscript{ST}), Universit{\"a}t Ulm, Albert-Einstein-Allee 11, D-89069 Ulm, Germany}}
\newcommand{\affTUDa}{\address{Technische Universit{\"a}t Darmstadt, Fachbereich Physik, Institut f{\"u}r Angewandte Physik, Schlossgartenstr. 7, D-64289 Darmstadt, Germany}}

\newcommand{\orcid}[1]{\href{https://orcid.org/#1}{\includegraphics[width=7pt]{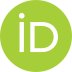}}}

\begin{document}

\title[Light propagation and atom interferometry in gravity and dilaton fields]{Light propagation and atom interferometry in gravity and dilaton fields}
\collaboration{This article has been published in \href{https://doi.org/10.1103/PhysRevD.105.084065}{Physical Review D \textbf{105}, 084065 [2022]}}

\author{Fabio Di Pumpo\,\orcid{0000-0002-6304-6183}}
\email{fabio.di-pumpo@uni-ulm.de}
\email{fabio.di-pumpo@gmx.de}
\affULM

\author{Alexander Friedrich\,\orcid{0000-0003-0588-1989}}
\affULM

\author{Andreas Geyer\,\orcid{0000-0001-8213-8350}}
\affULM

\author{Christian Ufrecht\,\orcid{0000-0003-4314-9609}}
\affULM

\author{Enno Giese\,\orcid{0000-0002-1126-6352}\,}
\affTUDa 
\affHAN

\begin{abstract}
\noindent
Dark matter or violations of the Einstein equivalence principle influence the motion of atoms, their internal states as well as electromagnetic fields, thus causing a signature in the signal of atomic detectors.
To model such new physics, we introduce dilaton fields and study the modified propagation of light used to manipulate atoms in light-pulse atom interferometers.
Their interference signal is dominated by the matter's coupling to gravity and the dilaton.
Even though the electromagnetic field contributes to the phase, no additional dilaton-dependent effect can be observed.
However, the light's propagation in gravity enters via a modified momentum transfer and its finite speed. 
For illustration, we discuss effects from light propagation and the dilaton on different atom-interferometric setups, including gradiometers, equivalence principle tests, and dark matter detection.
\end{abstract}

\maketitle

\section{Introduction}
Violations~\cite{Haugan1979,Damour1996,Will2014,Hees2018} of the Einstein equivalence principle (EEP)~\cite{Einstein1905,Einstein1907,DiCasola2015} and the existence~\cite{Damour1990,Boehm2004} of dark matter could be explained by scalar, light dilaton fields~\cite{Alves2000,Damour2010,Damour2012}.
They can be motivated by String Theory~\cite{Damour1994} and influence the propagation of light and matter, leading to deviations from established laws of physics.
While effects on matter have been thoroughly discussed~\cite{Damour1999,Arvanitaki2015,Geraci2016,Arvanitaki2018,Hees2018,ElNeaj2020,Roura2020,Roura2020b,Ufrecht2020,Badurina2021,DiPumpo2021,Zhao2021}, we focus on the electromagnetic field interacting with the dilaton and study the detection of dilaton fields by light-pulse atom interferometers~\cite{Kasevich1991,Cronin2009}.

While more advanced models are possible~\cite{Will2014}, the dilaton field already captures EEP violations and dark-matter effects, i.e., signatures of physics beyond the Standard Model and general relativity.
The dilaton field can be influenced by the presence of large masses like Earth, and thus induce violations of the universality of the gravitational redshift (UGR) and of free fall (UFF), as detected by EEP tests~\cite{Pound1960,Hafele1972a,Hafele1972b,Vessot1980,Godone1995,Fray2004,Chou2010,Bonnin2013,Schlippert2014,Ashby2018,Zhang2018,Delva2019,Touboul2019,Savalle2019,Asenbam2020,Takamoto2020,Bothwell2021}.
Additionally, an oscillating background field of cosmic origin can serve as a simple model for dark matter and gives rise to oscillating fundamental constants~\cite{Damour1999,Arvanitaki2015,Arvanitaki2018}, like the fine-structure constant.
Since the internal atomic structure depends on the value of these constants, atomic clocks~\cite{Moller1956,Katori2003,Nicholson2015,Brewer2019,Madjarov2019,Oelker2019} can be used to search for EPP violations~\cite{Damour1999,Hees2018} and dark matter~\cite{Arvanitaki2015,Derevianko2018,Wcislo2018}.
Similarly, in light-pulse atom interferometers~\cite{Kasevich1991,Cronin2009,Kleinert2015} matter propagates within dilaton fields and consequently such devices are susceptible~\cite{Tino2021} to both EPP violations~\cite{Giulini2012,Roura2020,Roura2020b,Ufrecht2020,DiPumpo2021} and dark matter~\cite{Geraci2016,Arvanitaki2018,ElNeaj2020,Badurina2021}.
However, with light pulses being an essential tool to manipulate the atoms, their modified behavior in gravity~\cite{Einstein1911,Tsagas2004,Bruschi2014,Cabral2017,Exirifard2021,Exirifard2021b,Mavrogiannis2021} and dilaton fields has to be taken into account for a consistent description of such experiments.

In this article we show that the propagation of light in dilaton fields has to lowest order no influence on the signal of an atom interferometer, while gravity leads to a modified momentum transfer.
Indeed, the dominant dilaton contribution to the interference pattern originates from the propagation of matter in the dilaton field.
To showcase the combined influence of light propagation and dilaton on atom interferometers, we study different schemes including gradiometers, EEP tests, and dark-matter detectors.
For the latter, we consider the spatial dependence of the dilaton field and discuss its consequences, which was not taken into account in previous treatments~\cite{Geraci2016,Arvanitaki2018,ElNeaj2020,Badurina2021}.

\section{Modified Maxwell equations}
We consider a classical dilaton field coupling linearly to all particles and forces in the Standard Model~\cite{Damour1994,Alves2000,Damour2010,Damour2012}.
The modified Lagrangian density for the electromagnetic sector is
\begin{align}
	\begin{split}
		\label{eq:DilIntLagFree}
		\mathcal{L}_\text{EM}=-\frac{\sqrt{-\mathfrak{g}}}{4\mu_0}(1-\varrho d_e)F_{\mu\nu}F^{\mu\nu}, 
	\end{split}
\end{align}
which includes a linear coupling between the dilaton field $\varrho$ and the electromagnetic field strength tensor $F_{\mu\nu}=\nabla_{\mu}A_\nu-\nabla_{\nu}A_\mu=\partial_{\mu}A_\nu-\partial_{\nu}A_\mu$ with coupling constant $d_e$ and vacuum permeability $\mu_0$.
Here, we introduced the determinant $\mathfrak{g}$ of the metric tensor, the covariant derivative $\nabla_\mu$, the partial derivative $\partial_\mu$, as well as the four-vector potential $A_\mu$.
The variation of this Lagrangian density with respect to $A_{\nu}$ leads to the Euler-Lagrange equation
\begin{align}
	\begin{split}
		\frac{\partial\mathcal{L}_\text{EM}}{\partial A_{\nu}}=\nabla_{\mu}\left(\frac{\partial\mathcal{L}_\text{EM}}{\partial(\nabla_{\mu}A_{\nu})}\right),
	\end{split}
\end{align}
which gives rise to modified Maxwell equations. 
Making use of the identity $\nabla_{\mu}\sqrt{-\mathfrak{g}}=0$, we find
\begin{align}
	\begin{split}
		\nabla_{\mu}(1-\varrho d_e) F^{\mu\nu}=0
	\end{split}
\end{align}
valid in the absence of four-currents, i.e., $\partial\mathcal{L}_\text{EM}/\partial A_{\nu}=0$~\footnote{The same result can be obtained from a similar variation with respect to $\partial_\mu A_\nu$ instead of $\nabla_{\mu}A_\nu$}.
Since the dilaton does not alter the metric tensor to lowest order (see Appendix~\ref{sec:App}), it can still be used to transform co- into contravariant quantities and vice versa.
Furthermore, adding a gauge field to the vector potential leaves the field strength tensor invariant and gauge freedom allows choosing the modified Lorenz gauge $\nabla_{\mu}(1-\varrho d_e)A^{\mu}=0$.
Together with the identity $\nabla_{\mu}\varrho=\partial_{\mu}\varrho$ and the Ricci tensor $R^\nu_\sigma$, we find the Maxwell equations
\begin{align}
	\begin{split}
		\nabla_{\mu}\nabla^{\mu}A^\nu=R^\nu_\sigma A^\sigma+(\partial_{\mu}\varrho)d_e\nabla^{\mu}A^\nu+	\nabla_\mu(\partial^\nu\varrho)d_e A^\mu, 
	\end{split}
\end{align}
neglecting orders of $d^2_e$. 
In the following, we consider a metric $\mathfrak{g}_{\mu\nu}=\eta_{\mu\nu}+\delta^0_\mu\delta^0_\nu 2 g z /c^2$ with gravitational acceleration $g$ pointing in $z$ direction and with  $R^\nu_\sigma=0$. 
Here, $\delta_\mu^{\nu}$ is the Kronecker delta and $\eta_{\mu\nu}$ the Minkowski metric. 
We keep only terms of order $g z/c^2$ as well as $d_e$, omitting higher contributions in the slowly-varying and weak-field limit for the source mass-energy density.
In Appendix~\ref{sec:App}, the metric is derived together with the solution for the dilaton field under the assumption of a source mass-energy density homogeneously distributed over an infinite plane.

We rely on the geometrical-optics ansatz~\cite{Dolan2018,Bartelmann2019,Linnemann2021}
\begin{align}
	\begin{split}
	A^{\nu}=\left(a^\nu+\epsilon b^\nu\right)\ee^{-\ii \Phi/\epsilon}+\mathcal{O}\big(\epsilon^2\big)
	\end{split}
\end{align}
to obtain propagating-beam solutions to the Maxwell equations.
In this approximation we assume that $\Phi$ is of order unity so that the factor $\exp{(-\ii\Phi/\epsilon)}$ varies fast compared to the amplitude $a^\nu$ and is common for all four components of the vector potential.
By combining the radiation wavelength $\lambda$ and the typical length scale $L$ of light propagation, we form a small parameter $\epsilon=\lambda/L$ for the perturbative solution of the Maxwell equations~\footnote{For more general situations where also a curvature is included into the metric, we generalize the definition to $\epsilon=\lambda/\min{(L,\mathcal{C})}$. Here, $\mathcal{C}\sim 1/\sqrt{\max{\left|R_{\mu\nu\kappa\sigma}\right|}}$ is the curvature radius scale.}.
For red laser light around $\lambda \sim 700$\,nm and for distances of $L\sim 1$\,mm, both typical scales for atom interferometry, we find $\epsilon= 7\cdot 10^{-4}$.
In more ambitious atom-interferometric setups, larger distances or different wavelengths give rise to an even more suitable regime.
After applying the perturbative expansion the factor $\epsilon$ can be absorbed into a redefinition of the wave vector.

\begin{center}
\begin{figure}[h]
	\includegraphics[width=1\columnwidth]{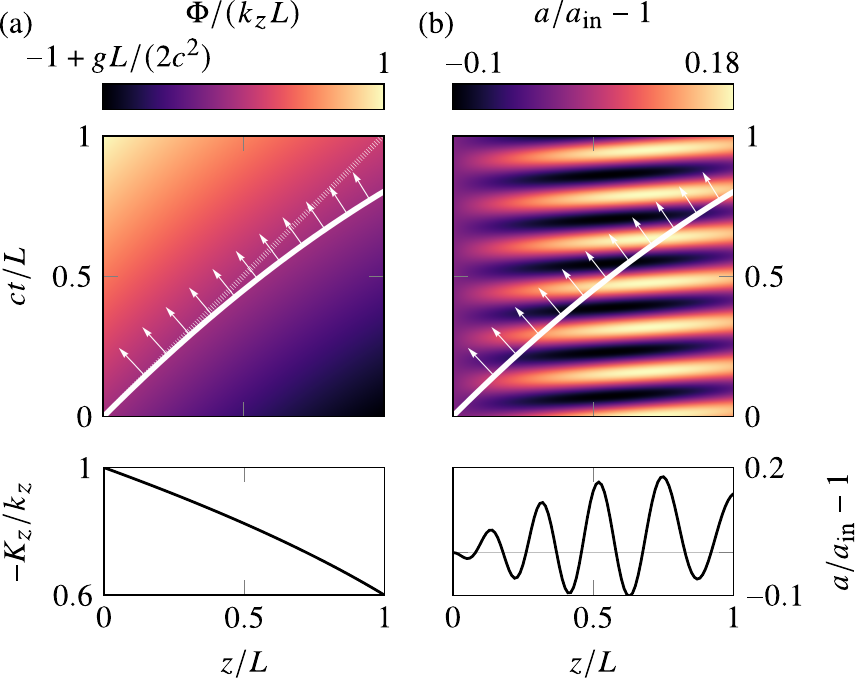}
	\caption{Effects of gravity and dilaton on the propagation of the electromagnetic field:
	The density plot (a) displays the scaled phase $\Phi/(k_z L)$ in a spacetime diagram, where $L$ is a typical length scale of the experiment.
	A light cone defined by a plane of constant phase (white curve) is deflected by gravity compared to the gravity-free case (dashed white line).
	The four-wave vector $K_\mu$ is the gradient of the phase and therefore orthogonal to the light cone.
	The two relevant components  $(K_0,K_z)$ are denoted by white arrows.
	As expected, the dilaton has no influence on the phase and the wave vector.
	A plot of the decreasing $z$ component $-K_z/k_z$ along the light cone is shown on the bottom.
	The amplitude $a/a_\text{in}-1$ is displayed by the density plot in panel (b) in a spacetime diagram.
	Again, we include the light cone (white curve) and the four-wave vector $K_\mu$ (white arrows) for reference.
	Although the phase and wave vector are not influenced by the dilaton, the underlying amplitude is modified by gravity and by the dilaton field with its oscillating background term.
	To highlight this effect, a cut of the oscillating amplitude along the light cone is shown on the bottom.
	For all plots, we use the exaggerated values $g L /(2c^2)= 0.2$, $d_e \bar{\beta}_\text{S}=0.3$, $d_e \bar{\varrho}_0= 0.02$, $k_\varrho L = 5$, $\omega_\varrho L /c = 40$, $\phi_\varrho = 0$, and $\boldsymbol{q}=0$ to enhance the visibility of the effect.}
\label{fig:PhaseAndAmplitude}
\end{figure}
\end{center}

When we define the wave vector $K_\mu=\partial_\mu \Phi$ and insert the ansatz into the Maxwell equations we obtain to leading order in $\epsilon^{-2}$ the condition $K_\mu K^\mu=0$.
This expression directly implies the eikonal equation
\begin{align}
	\begin{split}
	\mathfrak{g}^{\mu \nu}\partial_\mu \Phi\partial_\nu\Phi=0.
	\end{split}
\end{align}
As a consequence, the phase is independent of the dilaton field $\varrho$.
Since the metric only depends on the $z$ direction, we make the ansatz $\Phi=ck_0t-k_x x-k_y y-\kappa(z)$ to separate variables with $\kappa(0)=0$ and constants $k_0$, $k_x$ and $k_y$.
In the following, we define $\boldsymbol{q}=(k_x,k_y)^{\text{T}}$ and $\boldsymbol{r}=\left(x,y\right)^{\text{T}}$.
Together with the eikonal equation, we find $\kappa(z)=k_z\big[1-gk_0^2z/(2c^2k^2_z)\big]z$ for real constants $k_z$ and $k_0^2=\boldsymbol{q}^2+k_z^2$. 
We observe that the phase of the electromagnetic field in $z$ direction, and thus also the wave vector, are modified by gravity via the additional term $gk_0^2z^2/(2c^2k_z)$.
The phase $\Phi$ is shown in the spacetime diagram of Fig.~\ref{fig:PhaseAndAmplitude}\hyperref[fig:PhaseAndAmplitude]{(a)}, where we also plot the wave vector $K_\mu$ as a vector field along a light cone of constant phase.
The light cone is deflected by gravity compared to the flat-spacetime case, but experiences no effect caused by the dilaton.
The figure also shows the change of the $z$ component of the wave vector along the light cone.

Conventionally, one separates $a^\nu=a e^\nu$ into real amplitude $a=\sqrt{a^*_\nu a^{\nu}}$ and complex polarization $e^\nu$ with $e^*_\nu e^{\nu}=1$.
To order $\epsilon^{-1}$ we find from the gauge condition that $K_\mu e^\mu=0$, which implies an orthogonality between wave vector and polarization.
With this result, we find from the Maxwell equations to order $\epsilon^{-1}$ the equations of motion
\begin{align}
        \begin{split}
		&K^\mu\partial_\mu a+\frac{a}{2}\left[\nabla_\mu-\big(\partial_\mu\varrho\big)d_e\right]K^\mu=0,\\
		&K^\mu\nabla_\mu e^\nu=0
		\end{split}
\end{align}
for amplitude $a$ and polarization $e^\nu$, respectively.
Therefore, the dilaton modifies only the scalar amplitude but not the polarization.
To solve these equations, we use the explicit form of the Christoffel symbols included in the covariant derivative as well as the specific form of the dilaton field.
For the metric linear in $z$ we find $\Gamma^\nu_{0\lambda}=g(\delta^\nu_0 \delta^z_\lambda+\delta^\nu_z \delta^0_\lambda)/c^2$ and $\Gamma^\nu_{j\lambda}=\delta^\nu_0\delta^z_j \delta^0_\lambda g/c^2$.
Moreover, the dilaton field $\varrho=\bar\varrho_0\cos{\left(\omega_\varrho t-k_\varrho z+\phi_\varrho\right)}+\bar\beta_\text{S}g z/c^2$ arises for a homogeneously distributed mass-energy density, see Appendix~\ref{sec:App}.
Here, $\bar\beta_\text{S}$ corresponds to the linear expansion coefficient of the source mass $m_\text{S}$ of the gravitational field around its Standard-Model value, while $\bar\varrho_0$ is a perturbative amplitude.
The plane-wave part of the dilaton field has a frequency $\omega_\varrho$, an initial phase $\phi_\varrho$, and a wave number $k_\varrho$.
Thus, the dilaton field is a superposition of a contribution caused by the source mass and an oscillating background part.
As a consequence of these considerations, the components $e^x$ and $e^y$ of the polarization in $x$ and $y$ direction are constant and given by their initial values.
In contrast, for the other directions of the polarization we obtain 
\begin{align}
	\begin{split}
		&e^0(z)=e^0_\text{in}\left[1-\frac{g z}{c^2}\right]-e^z_\text{in}\frac{g z}{c^2}\frac{k_0}{k_z}, \\
		&e^z(z)=e^z_\text{in}-e^0_\text{in}\frac{g z}{c^2}\frac{k_0}{k_z},
	\end{split}
\end{align}
where $e^\nu_\text{in}$ denotes the initial value of each component.
Moreover, to lowest order in all perturbative parameters, neglecting cross terms, the amplitude
\begin{align}
\begin{split}
    a(z,t)=a_\text{in}&\left[1+\frac{g z}{2 c^2}\left(\frac{\boldsymbol{q}^2}{k^2_z}+d_e\bar\beta_\text{S}\right) +\left(1-\frac{\omega_\varrho}{ck_\varrho}\frac{k_0}{k_z}\right)\right.\\
    &\left.\times d_e\bar\varrho_0\sin{\left(\frac{k_\varrho z}{2}\right)}\sin{\left(\omega_\varrho t-\frac{k_\varrho z}{2}+\phi_\varrho\right)}\right]
\end{split}
\end{align}
with real initial value $a_\text{in}$ depends on time and position.
Since the square of the amplitude $a(z,t)$ is proportional to the energy density, we observe that it also depends on the dilaton and by that on time. 
This effect can be directly seen in Eq.~\eqref{eq:DilIntLagFree} from the modification of the Lagrangian density $\mathcal{L}_\text{EM}$ by the dilaton field.
For $k_0=k_z$ and vanishing dilaton mass $m_\varrho$, the time-dependent dilaton modification to the amplitude ceases to exist.
As expected from the $z$ dependence of the metric, the polarization only changes in vertical direction (and only via the metric tensor, not influenced by the dilaton).
In Fig.~\ref{fig:PhaseAndAmplitude}\hyperref[fig:PhaseAndAmplitude]{(b)} we plot the amplitude of the electric field in a spacetime diagram.
The light cone and wave vector $K_\mu$ from Fig.~\ref{fig:PhaseAndAmplitude}\hyperref[fig:PhaseAndAmplitude]{(a)} are included for reference.
We observe that the field experiences different amplitudes for different points in spacetime, depending on gravity and the underlying dilaton field.
Moreover, we illustrate this effect with the cut of the amplitude along the light cone shown on the bottom.

Since the electric field is obtained by taking the derivative of the vector potential, its phase corresponds to the same one as of $A^\nu$.
This fact implies that the frequency of the electric field $c k_0$ is independent of the position within the chosen frame of reference, and displays no gravitational modification.
In particular, this frequency experiences to leading order no modifications from the dilaton, which are only included in the amplitude of the electric field.
For illustration, we imagine two fixed atomic clocks at different heights that probe for violations of the gravitational redshift~\cite{Pound1960,Chou2010,Takamoto2020,Bothwell2021}.
The initialization and readout of these clocks is performed with light pulses.
During the interaction, the phase of the electric field is imprinted onto the atoms.
As a central result we find that this phase is independent of the dilaton field, and possible violations of the gravitational redshift measured by these clocks arise due to the coupling of matter to gravity and the dilaton field, and not from the modified phase of the electric field itself.

Conventionally, clocks are based on recoilless internal transitions~\cite{Alden2014,Bothwell2021} during the initialization and readout process.
However, for general pulses, the momentum of light carried by the wave vector cannot be neglected, and experiences a gravitational modification. 
The absorption and emission process causes a recoil of the first-quantized atom with center-of-mass position operator $(\hat{\boldsymbol{r}},\hat{z})^{\text{T}}$.
It is usually encoded~\cite{Kleinert2015} in the mode function $\exp{\big(\pm \ii\left[\boldsymbol{q}\hat{\boldsymbol{r}}+k_z\hat{z}\right]\big)}$ that gives rise to the momentum displacement $\pm (\hbar\boldsymbol{q},\hbar k_z )^\text{T}$.
However, this displacement operator is modified to $\exp{\big(\pm \ii [\boldsymbol{q}\hat{\boldsymbol{r}}+k_z  \hat{z}-gk_0^2\hat{z}^2/(2c^2k_z) ]\big)}$ in the presence of gravity.
Since setups like atom interferometers are built on this momentum transfer, there is an effect due to the altered wave vector.

\section{Atom-interferometric experiments}
\begin{figure}
	\centering
	\includegraphics{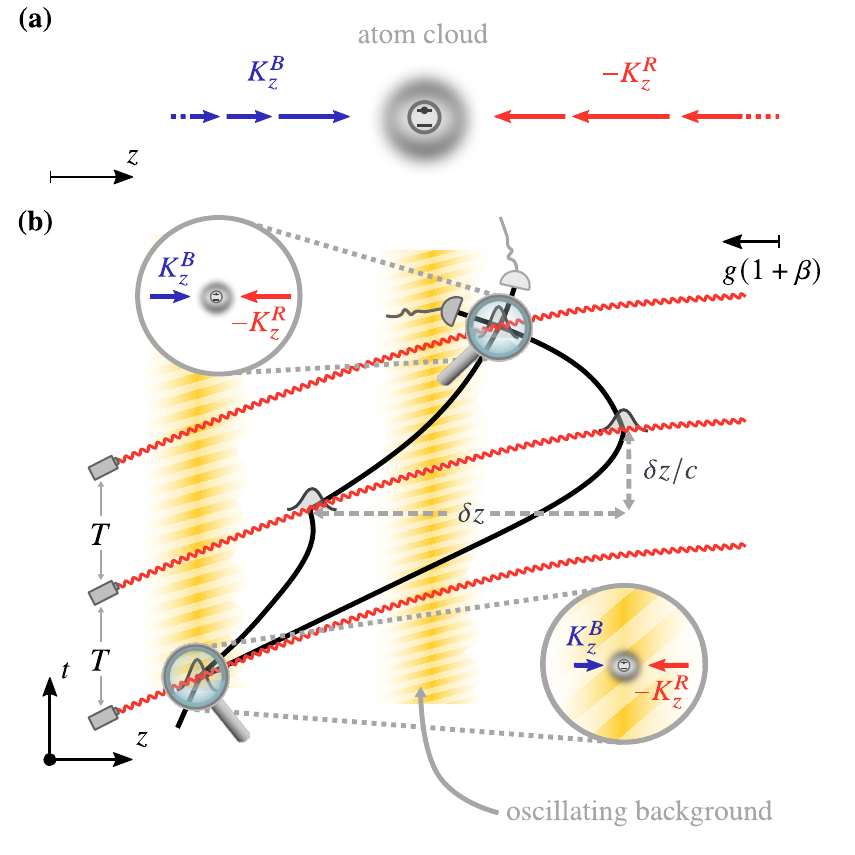}
	\caption{Effects of perturbations on atom interferometers:
	In (a) we depict diffraction from two counterpropagating light fields with different colors.
	We illustrate the opposite contractions of the blue and red wave vectors $K^\text{B}_z$ and $K^\text{R}_z$, originating from the propagation in different directions.
	As a consequence, they lead to a modified effective wave vector $k\big[1-gz/(2c^2)\big]$.
	A complete Mach-Zehnder sequence is shown in (b), where an initial $\pi/2$ pulse creates a spatial superposition, redirected by a $\pi$ pulse after time $T$, and brought to interference by a final $\pi/2$ pulse after another time interval $T$.
	Here, we draw the unperturbed atomic trajectories but highlight the effects due to perturbations.
	These perturbations include the modified wave vector of the effective pulse represented by the curved red light cones due to $k\big[1-gz/(2c^2)\big]$.
	The inset lenses illustrate that the diffracting wave vectors change for different pulses. 
	Moreover, the finite speed of light leads to a delay of order $\delta z/c$ between the diffraction of the lower and the upper branch for the $\pi$ pulse.
	Finally, we also include the coupling of the atom's mass $m$ to the dilaton field by two aspects: the mass-dependent gravitational acceleration $\beta g$ and the oscillating background represented by the shading between white and yellow.}
\label{fig:MZI}
\end{figure}

For the following calculations, we restrict ourselves to unidimensional interferometer setups in $z$ direction and thus choose $\boldsymbol{q}=0$ and $k_0=k_z$.
Conventionally, atom interferometers are operated via two-photon processes like Raman~\cite{Kasevich1991,Hartmann2020} or Bragg~\cite{Torii2000,Giese2015} diffraction, but recently single-photon diffraction has also gained some attention~\cite{Hu2017,Rudolph2020}.
Two-photon transitions usually involve a red-detuned laser with $k_z^\text{R}$ and a counterpropagating blue-detuned laser with $k_z^\text{B}$.
Both wave vectors are defined by their respective value at the surface of the source mass, i.e., $z=0$.
They give rise to the differential phase of both light fields $\Phi_B-\Phi_R=\Delta \omega t-kz\big[1-gz/(2c^2)\big]$.
Here, $\Delta\omega=c(k_z^\text{B}-k_z^\text{R})$ is the frequency difference and is proportional to the transferred energy during resonant diffraction, whereas $k=k_z^\text{B}+k_z^\text{R}$ is the effective wave vector transferring momentum, which experiences a modification by the factor $1-gz/(2c^2)$.
Thus, the gravitationally modified wave vector for a two-photon process has the same form as the one from a single-photon process, even though $\Delta\omega$ and $k$ can be tuned independently.
Figure~\ref{fig:MZI}\hyperref[fig:MZI]{(a)} illustrates the impact of the modified wave vector on the red and blue lasers involved in a two-photon process via arrows with different lengths.
We neglect the motion of the atom during the pulse, which corresponds to a vanishingly small pulse duration compared to the duration of the interferometer.
In this case, the interferometer can be described by a series of instantaneous momentum transfers with the effective potential 
\begin{align}
\label{eq:kick-potential}
    \hat{V}=-\hbar\sum_n{k_n\hat{z}\left[1-\frac{g\hat{z}}{2c^2}\right]\delta(t-t_n)}= \hat{V}_0+\hat{\mathcal{V}},
\end{align} 
where the momentum transfer $\hbar k_n$ of the $n$th light pulse at time $t_n$ is modified by gravity.
The time evolution with this potential directly modifies the displacement operator to the operator introduced in the preceding section.
Moreover, the second term in the brackets, defined as $\hat{\mathcal{V}}$, can be interpreted as a small perturbation to the momentum transfer without gravity included in $\hat{V}_0$. 
The dilaton-dependent amplitude of the vector potential only leads to a modified Rabi frequency and, therefore, does not contribute to the momentum transfer to first-order perturbation theory.

However, the dilaton indeed contributes to the phase of an atom interferometer~\cite{Geraci2016,Arvanitaki2018,ElNeaj2020,Roura2020,Roura2020b,Ufrecht2020,Badurina2021,DiPumpo2021,Zhao2021} because the complete Hamiltonian for a particle of mass $m$ and momentum $\hat{p}$ is
\begin{align}
\label{eq:Hamiltonian}
\begin{split}
    \hat{H}^{(\sigma)}=&m c^2 +\frac{\hat{p}^2}{2m}+m g\hat z +\hat{V}^{(\sigma)}_0\\
    &+\hat{\mathcal{V}}^{(\sigma)}+m g\beta\hat z+ m c^2\varrho_0\cos{\left(\omega_\varrho t-k_\varrho\hat{z}+\phi_\varrho\right)}
\end{split}
\end{align}
and depends on the dilaton parameters $\beta=\bar\beta\bar\beta_\text{S}$ and $\varrho_0=\bar\beta\bar\varrho_0$.
It leads to a particle-dependent gravitational acceleration $g(1+\beta)$ for the center-of-mass motion between the pulses, introducing EPP violations~\cite{Damour1999,Giulini2012,Hees2018,Roura2020,Roura2020b,Ufrecht2020,DiPumpo2021,Fray2004,Zhang2018}.
The Hamiltonian also includes time-dependent modifications caused by the dilaton field that are independent of gravity, which can be connected to dark matter~\cite{Arvanitaki2015,Geraci2016,Arvanitaki2018,ElNeaj2020,Badurina2021}.
Here, we anticipated that an atom interferometer consists of at least two branches by introducing the superscript $\sigma = u, l$ for the upper and lower branch.
They are generated by branch-dependent momentum kicks $\hbar k^{(\sigma)}_n$ encoded in $\hat{V}^{(\sigma)}$ as a generalization of Eq.~\eqref{eq:kick-potential}.
The dilaton parameter $\bar\beta$ effectively corresponds to the linear expansion coefficient of the mass $m$ around its Standard-Model value.

Our goal is to identify the leading-order contributions of the dilaton field and the propagation of light, including its finite speed, to the signal of atom interferometers.
To keep focus, we neglect other effects in our description, such as relativistic kinetic corrections, possibly in the same order of magnitude.

To compute phases in dilaton gravity, we consider the Bragg-Mach-Zehnder~\cite{Torii2000} geometry shown in Fig.~\ref{fig:MZI}\hyperref[fig:MZI]{(b)}, where in contrast to Raman-based schemes the pulses induce no internal transitions.
The figure highlights the impact of the modified potential by curved light cones representing effective Bragg pulses, and arrows with different lengths at two interaction points transferring position-dependent momentum. 
Moreover, we also indicated the perturbations from finite speed of light, modified wave vectors, and dilaton fields described in more detail below. 
The scheme consists of an initial $\pi/2$ light pulse to create a spatial superposition, followed by a $\pi$ pulse to redirect the atomic trajectories, and finally another $\pi/2$ pulse to interfere both branches.
The pulses are separated by a time interval $T$.
For the Bragg-Mach-Zehnder scheme we calculate the phase via a perturbative formalism~\cite{Ufrecht2019,Ufrecht2020b}:
We split the Hamiltonian into an unperturbed part $\hat{H}^{(\sigma)}_0$, defined by the first line of Eq.~\eqref{eq:Hamiltonian}, and a perturbation $\hat{\mathcal{H}}^{(\sigma)}$ given by the second line.
Moreover, we consider the classical trajectories generated by a classical counterpart of $\hat{H}^{(\sigma)}_0$.
These unperturbed trajectories are inserted into the perturbing classical counterpart of $\hat{\mathcal{H}}^{(\sigma)}$ and integrated along the upper and the lower branch.
For unperturbed interferometers closed in phase space this procedure leads to a phase
\begin{align}
\label{eq:general_phase}
    \varphi=\varphi_0-\frac{1}{\hbar}\int\text{d}t~\left(\mathcal{H}^{(u)}-\mathcal{H}^{(l)}\right)+\varphi_\text{FSL}.
\end{align} 
Here, $\varphi_0=-kgT^2$ arises from the unperturbed Hamiltonian $\hat{H}^{(\sigma)}_0$~\cite{Kleinert2015}.
Cross terms between perturbations are of higher order and are, consequently, neglected.
Since we are interested in relativistic effects due to light propagation, we also include~\cite{Dimopoulos2008,Tan2016,Tan2017} the finite-speed-of-light contribution $\varphi_\text{FSL}$.
This additional term $\varphi_\text{FSL}=-3 k g T^2 v_T/c$ depends on the velocity $v_T=v_0-g T+\hbar k/m$ on the upper branch at the central pulse, with initial velocity $v_0$ at the first pulse.
Although $\varphi_\text{FSL}$ has been derived before~\cite{Dimopoulos2008,Tan2016,Tan2017}, we analyze its effect in different situations. 

\subsection{Gravitational effects including the dilaton}
In this section we analyze the effect of gravity including the dilaton field on atom interferometry both via light propagation and the coupling of the atomic mass to the dilaton. 
To highlight the gravitational contributions, we set $\varrho_0=0$, i.\,e., we consider only the influence of the gravitational source mass on the dilaton field and neglect the background oscillation.
To lowest order in all perturbative parameters we obtain the phase
\begin{align}
\label{eq:SinglePhase}
	\begin{split}
		-\frac{\varphi}{k g T^2}=&1+\beta+3\frac{v_T}{c}+ \frac{v_0v_T}{c^2}- \frac{g z_T}{c^2}\\
		&- \frac{g T}{c^2} \left( v_0+ \frac{\hbar k}{2 m} - g T\right)+ \frac{g^2T^2}{4c^2}
	\end{split}
\end{align}
apart from a trivial laser contribution.
Here, we introduced the atom's initial position $z_0$, as well as the atomic position $z_T=z_0+v_0T-gT^2/2+\hbar k T/m$ on the upper branch during the central pulse.
The first term represents the conventional gravimeter phase~\cite{Kasevich1991,Kleinert2015}.
The second term corresponds to the leading dilaton violation of EEP~\cite{Fray2004,Bonnin2013,Schlippert2014,Zhang2018,Asenbam2020}.
The third term stems from finite speed of light~\cite{Dimopoulos2008,Tan2016,Tan2017}.
All remaining terms result from the modified wave vector.
They resemble phases caused by gravity gradients~\cite{Kleinert2015}, since the perturbative potential in Eq.~\eqref{eq:kick-potential} is quadratic in the atom's position.
In principle, all additional contributions in Eq.~\eqref{eq:SinglePhase} limit the sensitivity of gravimeters, but are usually negligibly small compared to other deleterious effects.
However, in the following we discuss their effect in differential measurements where other noise sources are suppressed.

\subsubsection{Atom-interferometric gradiometry}
Atom-interferometric gradiometers~\cite{Snadden1998,McGuirk2002,Abe2021} can be built from two identical interferometers, each with individual (but possibly phase-locked) lasers, vertically separated by a length $\ell$.
We assign an index $j=1,2$ to the upper and lower interferometer, which allows us to infer the differential phase $\delta\varphi=\varphi_1-\varphi_2$ between both signals.
Each phase is associated with a local linear acceleration $g_j$, with mean acceleration $g=(g_1+g_2)/2$ and difference $\delta g=g_1-g_2$.
They capture the change of the gravitational field due to higher gravitational multipoles between the atom interferometers and correspond to two different expansion points of the metric.
However, the treatment neglects local gravity gradients on the scale of the extension of the individual interferometer~\cite{Roura2014,DAmico2017,Ufrecht2021}.
Since effects from the modified wave vector scale with $c^{-2}$, we also neglect its cross terms with $\delta g$.
If we prepare equal initial velocities, we find the differential phase
\begin{align}
	\begin{split}
		\frac{\delta\varphi}{kgT^2}=-\frac{\delta g}{g}\left(1+\beta+3\frac{v_T-gT}{c}\right)+\frac{g\ell}{c^2}
	\end{split}
\end{align}
between both interferometers.
The contribution from the modified wave vector grows with increasing spatial separation and is of order $g\ell/c^2\cong10^{-13}$ for planned separations of up to $10^3\,$m for Earth-based setups~\cite{Abe2021}.
Even though such devices are mainly limited by gravity-gradient noise, this phase contribution has to be taken into account for a thorough analysis.
At the same time, it offers an atom-interferometric verification of the influence of gravity on light, so far primarily observed in an astronomical or cosmological context.
Similarly, finite speed of light sets limits on the accuracy of gradiometers and crucially depends on the velocity of the atoms at the central pulse.

\subsubsection{EEP tests with atom interferometry}
Atom-interferometric tests of EEP~\cite{Fray2004,Bonnin2013,Schlippert2014,Zhang2018,Asenbam2020,Roura2020,Roura2020b,Bothwell2021,Ufrecht2020,DiPumpo2021} involve differential phases between two species or atomic levels, labeled by the index $j=a,b$.
For two masses $m_a$ and $m_b$ we calculate the differential phase $\theta_k=\varphi_a/(|k_a| g T^2)-\varphi_b/(|k_b| g T^2)$, where $m_j=m+\lambda_j\Delta m/2$ with $\lambda_b=1$ and $\lambda_a=-1$. 
Here, $m = (m_a+m_b)/2$ is the mean mass.
Similarly, each atomic component is addressed by a different laser and therefore experiences a different momentum transfer $\hbar k_j = \hbar k+\lambda_j\hbar \Delta k/2$.
We also introduce different violation parameters $\beta_j$, as well as different initial values for position and velocity.
For $k_b,k_a>0$ we obtain
\begin{align}
\label{eq:DiffPhase}
	\begin{split}
		\theta_k=&\Delta\beta+3\frac{\Delta v_0}{c}-\frac{g}{c^2}\left(\Delta z_0+3\Delta v_0T\right)+2\frac{v_0\Delta v_0}{c^2} \\
		&+ \chi  \frac{v_\text{r}}{c}\left(3-\frac{3g T}{2c}+ \frac{v_0}{c} \right)\left(\frac{\Delta k}{k}-\frac{\Omega}{\omega_C}\right)\\
		&+\chi\frac{v_\text{r}\Delta v_0}{c^2}\left(1-\frac{\Delta k}{4k}\frac{\Omega}{\omega_C}\right)
	\end{split}
\end{align}
with the violation parameter $\Delta\beta=\beta_b-\beta_a$, which can be used to parametrize UFF tests~\cite{Fray2004,Bonnin2013,Schlippert2014,Zhang2018,Asenbam2020}.
Here, we introduce the average recoil velocity $v_\text{r} = \hbar k /m$ and different initial velocities $v_0+\lambda_j\Delta v_0/2$ and positions $z_0+\lambda_j\Delta z_0$.
The parameter $\chi=\big[1-\Omega^2/(2 \omega_C)^2\big]^{-1}$ quantifies the magnitude of the mass difference, where we defined $\Omega=\Delta m c^2/\hbar$ and the frequency $\omega_C=m c^2/\hbar$.

Quantum clock interferometry~\cite{Sinha2011,Zych2011,Rosi2017,Loriani2019,Ufrecht2020,Roura2020,Roura2020b,DiPumpo2021} is based on a superposition of two different internal states.
Even though their eigenenergies are associated with different masses~\cite{Lammerzahl1995,[][ and references therein.]Pikovski2017,Sonnleitner2018,Schwartz2019} through the relativistic mass-energy relation, the factor $\Omega/\omega_C\ll1$ acts perturbatively.
Such interferometers are often operated with magic Bragg diffraction~\cite{Katori2003}, which implies $\Delta k=0$.
Similarly, the differences $\Delta v_0$ and $\Delta z_0$ for two internal states are strongly perturbative and can be neglected when divided by the speed of light.
As a consequence, only the dilaton-dependent term $\Delta\beta$ is relevant for EEP tests via quantum clock interferometry.

For different atomic species their mass difference is of the order of the mean mass so that the arguments for quantum clock interferometry do not apply.
However, when performing conventional UFF tests with distinct atomic species~\cite{Fray2004,Bonnin2013,Schlippert2014,Asenbam2020} the momentum kick $k_j$ is reversed~\cite{McGuirk2002} for both species after each experimental run. 
This way, noise and other deleterious effects are removed from the signal.
Hence, we find the two-fold differential phase
\begin{align}
\label{eq:TwoFoldPhase}
	\begin{split}
		\frac{\theta_k-\theta_{-k}}{2}=\Delta\beta+3\frac{\Delta v_0}{c}+2\frac{v_0\Delta v_0}{c^2}-\frac{g\Delta z_0}{c^2}-3\frac{g\Delta v_0T}{c^2}
	\end{split}
\end{align}
leading to the cancellation of all effects linear in $k_j$.
Apart from the UFF violation parameter, only effects from different initial positions and velocities survive this $k$-reversal technique.
These remaining terms are dominated by the finite-speed-of-light effect $3\Delta v_0/c$.
The estimate for already performed UFF tests~\cite{Bonnin2013} was $\Delta v_0/c\cong 2\cdot10^{-11}$, which is close to the uncertainty of planned experiments~\cite{Schlippert2020}.
As such, the effect has to be subtracted for the analysis of the test.
To this end, the error $\delta(\Delta v_0)$ caused by the uncertainty of the differential initial velocity remains~\cite{Nobili2016}.
Assuming~\cite{Loriani2020} $1\,$\textmu m/s leads to $\delta(\Delta v_0)/c\cong3\cdot10^{-15}$.
As a consequence, the remaining effect from finite speed of light in UFF tests with two species can be removed from the signal up to a precision of $10^{-15}$ due to good control over the initial velocities.
The other terms in Eq.~\eqref{eq:TwoFoldPhase} proportional to $\Delta v_0$ are further suppressed by the inverse speed of light.
They can be neglected for state-of-the-art and near-future schemes.
Similarly, the term proportional to $\Delta z_0$ is also of no relevance for UFF tests so far.
Finally, we see that only the atom's propagation in the dilaton field induces phases sensitive to UFF violations.
This observation can be tracked back to the fact that the dilaton only modifies the Rabi frequency of the diffracting pulse but not the momentum transfer.
Consequently, the propagation of light in dilaton gravity does not contribute to these violations.

\subsection{Effects of oscillating dilaton field}
In this section, we consider the coupling of the time-dependent part of the dilaton field, not caused by the source mass, to an atom interferometer, as for example used as a model in the context of dark-matter detection~\cite{Geraci2016}.
Similar to gradiometry, the detection scheme~\cite{Arvanitaki2018,ElNeaj2020,Badurina2021} is based on differential phases $\delta\varphi=\varphi_1-\varphi_2$ between two identical atom interferometers operated at different locations.
For example, in proposed space-based gravitational-wave~\cite{Dimopoulos2009,Hogan2016} and dark-matter detection schemes~\cite{ElNeaj2020} two atom interferometers are placed at distances kilometers apart from each other.
Due to microgravity, we have $g=0$.
This choice naturally leads to a cancellation of all finite-speed-of-light effects that arise within each interferometer, as wells as all effects from the modified wave vector, since they are proportional to the gravitational acceleration.
We calculate the phase of one atom interferometer similar to Eq.~\eqref{eq:SinglePhase} but with $g=0$, and solely include the plane-wave part of the dilaton field.
However, the interferometers interact with the light pulses at locations separated by $\ell$, which implies an offset $\ell/c$ between the start of the interferometers.
By taking the different initial times into account and assuming the same initial velocity $v_0$, we find the differential phase
\begin{align}
	\begin{split}
		\frac{\delta\varphi}{(ckT)^2}&=-2\varrho_0\frac{k_\varrho}{k} 
		\cos{\left(\omega_\varrho T -k_\varrho \bar z_T + \frac{\omega_\varrho \ell}{2c}+\phi_\varrho\right)} \\
		&\times \sin{\left( \frac{\omega_\varrho \ell}{2c} \left[1-\frac{ck_\varrho}{\omega_\varrho}\right]\right)}
		\operatorname{sinc}{\left(\frac{\omega_\varrho T}{2}\left[1-\frac{v_0 k_\varrho}{\omega_\varrho}\right]\right)}  \\
		&\times\operatorname{sinc}{\left(\frac{\omega_\varrho T}{2}\left[1-\frac{v_T k_\varrho}{\omega_\varrho}\right]\right)}
	\end{split}
\end{align}
with average position $\bar z_T=\ell+v_0 T+\hbar k T/(2m)$ and velocity $v_T=v_0+\hbar k/m$ on the upper branch at the central pulse of each interferometer.
Such schemes measure the different couplings of the dilaton to the atom's mass between the two interferometers.
These different couplings are caused by the oscillations of the background field. 
Since the initial phase $\phi_\varrho$ of the dilaton field is unknown, we find the amplitude of the fluctuations induced by the dilaton field and measured by a continuous operation of the detector from $\varphi_\text{SA}= \left(\int_0^{2\pi}{\mathrm{d}\phi_\varrho~\delta\varphi^2/\pi}\right)^{1/2}$.
It corresponds to the standard deviation of the phase signal.
The amplitude takes the form
\begin{align}
	\begin{split}
		\frac{\varphi_\text{SA}}{(ckT)^2}&=2\varrho_0\frac{k_\varrho}{k} 
		\sin{\left( \frac{\omega_\varrho \ell}{2c} \left[1-\frac{ck_\varrho}{\omega_\varrho}\right]\right)} \\
		&\times\operatorname{sinc}{\left(\frac{\omega_\varrho T}{2}\left[1-\frac{v_0 k_\varrho}{\omega_\varrho}\right]\right)}\operatorname{sinc}{\left(\frac{\omega_\varrho T}{2}\left[1-\frac{v_T k_\varrho}{\omega_\varrho}\right]\right)},
	\end{split}
\end{align}
which shows that the influence of the oscillating background can indeed be resolved in such differential setups.
We emphasize that for ceasing dilaton mass $m_\varrho$, i.e., $\omega_\varrho = c k_\varrho$, the signal amplitude vanishes.
Note that the assumption of the dilaton's Compton wave length being much larger than the extension of the experiment (see Appendix~\ref{sec:App}) is not necessary for such spaceborne schemes, since the gravitational part of the dilaton field vanishes here.
Similarly, for vanishing dilaton recoil $\hbar k_\varrho$, the limit usually discussed in the context of dark-matter detection with atom interferometers~\cite{Geraci2016,Arvanitaki2018,ElNeaj2020,Badurina2021}, the signal amplitude also vanishes.
As such, our result differs from these previous ones, as we perform only first-order perturbation theory but include also the dilaton mass and the dilaton recoil in our Hamiltonian from Eq.~\eqref{eq:Hamiltonian}.
Although the laser propagates through the dilaton field, this propagation does not imprint dilaton-dependent terms on the signal.
In contrast, similar to the Earth-based schemes discussed before, it is the atom's mass which probes the oscillating background imprinting the phase.

\section{Discussion}
We derived dilaton-modified Maxwell equations from a first-principle Lagrangian approach, and solved them with a geometrical-optics ansatz.
The resulting phase of the electromagnetic field is not influenced by the dilaton to lowest order, in contrast to the amplitude which experiences a dilaton-dependent modification.
Nevertheless, gravity alters the wave vector of the electromagnetic field.
Consequently, no dilaton effects are imprinted by the light's phase in atom-optical experiments, although the momentum transfer on the atom is modified by gravity.
However, the dilaton couples to the mass-energy of the atom and, therefore, affects the time evolution of the atomic structure and center-of-mass motion.

To illustrate our results, we analyzed two scenarios for a light-pulse Mach-Zehnder atom interferometer:
a dilaton field solely sourced by Earth's mass and an oscillating background field of cosmic origin.
In the first case, we showed that for gradiometers the effects from finite speed of light and the gravitationally modified wave vector are the dominant contributions.
For EEP tests, however, only the UFF violation parameter survives for quantum clock experiments performed with a superposition of internal states.
Differential effects are negligible because the mass-energy difference of two such states is small and no colocation issues arise in the state preparation.
Contrarily, for ambitious UFF tests with different species and a $k$-reversal technique, additional effects from finite speed of light can become relevant.
We discussed the detection of dark matter in spaceborne experiments by probing an oscillating dilaton field at different locations in spacetime with two atom interferometers.
Already a spatial dependence of the dilaton leads to measurable effects, a fact not accounted for in most studies of atom-interferometric dark-matter detectors.

Hence, this article provides a discussion of the leading-order propagation of all constituents relevant for light-pulse atom interferometers in gravity and dilaton fields, i.\,e., the propagation of light and matter.
While the propagation of matter gives rise to dominant effects, corrections from the propagation of light in gravity has to be taken into account for future high-precision experiments.
Thus, our results constitute a framework for the analysis of atomic experiments where the propagation of light plays a crucial role, such as gravitational wave or dark matter detectors.
Consequently, they can be used to constrain parameters that describe extensions of the Standard Model or general relativity, as exemplified by the dilaton field.
The potential of atom interferometry for fundamental physics can be highlighted by future studies of different geometries beyond the examples discussed in our article.

\begin{acknowledgments}
We are grateful to W. P. Schleich for his stimulating input and continuing support.
We also thank W. G. Unruh, as well as the QUANTUS and INTENTAS teams for fruitful and interesting discussions.
The projects ``Metrology with interfering Unruh-DeWitt detectors'' (MIUnD) and ``Building composite particles from quantum field theory on dilaton gravity'' (BOnD) are funded by the Carl Zeiss Foundation (Carl-Zeiss-Stiftung).
The work of IQ\textsuperscript{ST} is financially supported by the Ministry of Science, Research and Art Baden-W\"urttemberg (Ministerium f\"ur Wissenschaft, Forschung und Kunst Baden-W\"urttemberg).
The QUANTUS and INTENTAS projects are supported by the German Aerospace Center (Deutsches Zentrum f\"ur Luft- und Raumfahrt, DLR) with funds provided by the Federal Ministry for Economic Affairs and Climate Action (Bundesministerium f\"ur Wirtschaft und Klimaschutz, BMWK) due to an enactment of the German Bundestag under Grants No. 50WM1956 (QUANTUS V), No. 50WM2250D-2250E (QUANTUS+), as well as No. 50WM2177-2178 (INTENTAS).
E.G. thanks the German Research Foundation (Deutsche Forschungsgemeinschaft, DFG) for a Mercator Fellowship within CRC 1227 (DQ-mat).
\end{acknowledgments}

\appendix

\section{Modified Einstein equations}
\label{sec:App}
To include violations of EPP and model possible dark matter, we consider an extension~\cite{Alves2000,Damour2010,Damour2012,Damour1994} of the Standard Model on curved spacetime through a linear coupling to a classical, light dilaton field $\varrho$.
As such, the Lagrangian density
\begin{align}
    \begin{split}
    \label{eq:Largangian}
    \mathcal{L}=\mathcal{L}_\text{EH}+\mathcal{L}_\text{EM}+\mathcal{L}_\text{mat}
    \end{split}
\end{align}
includes the modified electromagnetic part $\mathcal{L}_\text{EM}$ from Eq.~\eqref{eq:DilIntLagFree}.
It is complemented by the modified Einstein-Hilbert action
\begin{align}
    \begin{split}
    \mathcal{L}_\text{EH}=\sqrt{-\mathfrak{g}}\frac{c^4}{16\pi G}\left[R-2\mathfrak{g}^{\mu\nu}\partial_\mu\varrho\partial_\nu\varrho -\frac{2\varrho^2}{\lambda^2_\varrho}\right]
    \end{split}
\end{align}
with a kinetic term $(\nabla\varrho)^2$ of the dilaton, the Ricci scalar $R$, the (bare) Newtonian gravitational constant $G$, the determinant $\mathfrak{g}$ of the metric tensor $\mathfrak{g}_{\mu\nu}$, and a (small) dilaton mass $m_\varrho$.
Here, $\lambda_\varrho=\hbar/(cm_\varrho )$ is the reduced Compton wavelength of the dilaton.
In Eq.~\eqref{eq:Largangian}, $\mathcal{L}_\text{mat}$ denotes the Lagrangian density of the (dilaton-modified) matter sector, consisting of (quantum) test bodies as well as (classical) sources for the gravitational and dilaton fields.
From the Einstein-Hilbert action we find by variation the modified Einstein equations
\begin{subequations}
\begin{align}
		&R_{\mu\nu}=2\partial_\mu \varrho\partial_\nu \varrho+\mathfrak{g}_{\mu\nu}\frac{\varrho^2}{\lambda^2_\varrho}+\frac{8\pi G}{c^4}\left(T_{\mu\nu}-\frac{1}{2}T\mathfrak{g}_{\mu\nu}\right) \\
		&\nabla_\mu\partial^\mu \varrho=\frac{1}{\sqrt{-\mathfrak{g}}}\partial_\mu\sqrt{-\mathfrak{g}}\partial^\mu\varrho=-\frac{4\pi G}{c^4}\sigma+\frac{\varrho}{\lambda^2_\varrho}, \label{eq:EinsteinEqDilaton}
\end{align}
\end{subequations}
with Ricci tensor $R_{\mu\nu}$. 
Here, the energy-momentum tensor $T_{\mu\nu}=-(2/\sqrt{-\mathfrak{g}}) \delta(\sqrt{-\mathfrak{g}}\mathcal{L}_\text{mat})/\delta \mathfrak{g}^{\mu\nu}$ is obtained from a variation of $\mathfrak{g}^{\mu\nu}$, and the energy-momentum scalar $\sigma=(1/\sqrt{-\mathfrak{g}})\delta(\sqrt{-\mathfrak{g}}\mathcal{L}_\text{mat})/\delta \varrho$ from a variation of $\varrho$.
The quantity $T=T^{\mu}_\mu$ is the Laue scalar.
We neglect the influence of quantized matter-energy on $T_{\mu\nu}$ and $\sigma$, such as quantized test bodies or the electromagnetic field.

For a mass-energy distribution on an infinite, homogeneous plane at $z=0$, we find the energy-momentum tensor $T_{\mu\nu}=\delta^0_\mu\delta^0_\nu m_\text{S}(0)\delta(z) c^2/A$.
Here, $A$ is an area element, $m_\text{S}(0)$ is the Standard-Model value of the source mass, $\delta_\mu^{\nu}$ is the Kronecker delta, and $\delta(z)$ is the delta function.
Defining $g=4\pi G m_\text{S}(0)/A$ as the acceleration orthogonal to the plane, this tensor takes the form $T_{\mu\nu}=\delta^0_\mu\delta^0_\nu g\delta(z) c^2/(4\pi G)$. 
Similarly, we find the energy-momentum scalar $\sigma=-g\bar\beta_\text{S}\delta(z) c^2/(4\pi G)$, where $\bar\beta_\text{S}=\left[1/m_\text{S}(0)\right]\left.\partial m_\text{S}/\partial\varrho\right|_{\varrho=0}$ corresponds to the linear expansion coefficient for the source mass $m_\text{S}(\varrho)$ around its Standard-Model value $m_\text{S}(0)$.
Hence, we find from Eq.~\eqref{eq:EinsteinEqDilaton} that $\nabla_\mu\partial^\mu \varrho=\bar\beta_\text{S} g\delta(z)/c^2+\varrho/\lambda^2_\varrho$, which dictates boundary conditions for the dilaton field on the plane.
For regions outside the source mass-energy distribution $z>0$, we obtain modified vacuum equations
\begin{subequations}
    \begin{align}
			&R_{\mu\nu}=2\partial_\mu \varrho\partial_\nu \varrho+\mathfrak{g}_{\mu\nu}\varrho^2/\lambda^2_\varrho, \\
			&\partial_\mu\sqrt{-\mathfrak{g}}\partial^\mu\varrho=\sqrt{-\mathfrak{g}}\varrho/\lambda^2_\varrho. \label{eq:SecondVakuum}
    \end{align}
\end{subequations}

These equations are dominated by the general-relativistic equation $R_{\mu\nu}=0$ for a weak perturbing dilaton field.
Thus, we consider the metric tensor $\mathfrak{g}_{\mu\nu}$ without any dilaton contribution.
We then solve for the dilaton field $\varrho$, which is coupled to this metric tensor.
For the mass distribution considered above, the vacuum equation is solved by a metric of the form $\mathfrak{g}_{\mu\nu}=\eta_{\mu\nu}+\delta^0_\mu\delta^0_\nu 2 g z/c^2$, with the gravitational acceleration $g$ aligned in $z$ direction and the Minkowski metric $\eta_{\mu\nu}$.
From Eq.~\eqref{eq:SecondVakuum} we find 
\begin{align}
	\begin{split}
	\left(\partial^2_{0}-\partial^2_j\right)\varrho=\frac{g}{c^2}\left(2 z\partial^2_j+\partial_z\right)\varrho+\frac{\varrho}{\lambda^2_\varrho}
	\end{split}
\end{align}
to lowest order in $g z/c^2$.
Since we are interested in light propagation and atom interferometers in $z$ direction, we assume the dilaton field to depend only on $t$ and $z$.
Furthermore, we split the dilaton field $\varrho(t,z)=\varrho_\text{h}(t,z)+\varrho_g(z)$ into a homogeneous part $\varrho_\text{h}(t,z)$ and a gravitational perturbation $\varrho_g(z)$ of order $g z/c^2$, which depends solely on $z$.
For the homogeneous part we obtain the Klein-Gordon equation
\begin{align}
	\begin{split}
	\left(\partial^2_{0}-\partial^2_z\right)\varrho_\text{h}=\varrho_\text{h}/\lambda^2_\varrho,
	\end{split}
\end{align}
which is solved by $\varrho_\text{h}=\bar\varrho_0\cos{\left(\omega_\varrho t-k_\varrho z+\phi_\varrho\right)}$.
The dilaton has the (perturbative) amplitude $\bar\varrho_0$, the wave number $k_\varrho$, the frequency $\omega^2_\varrho= (ck_\varrho)^2 + (c/ \lambda_\varrho)^2$, and the initial phase $\phi_\varrho$.
With this solution, we find for the gravitational perturbation
\begin{align}
	\begin{split}
	-\partial^2_z\varrho_g&=\frac{g}{c^2}\left(2 z\partial^2_z+\partial_z\right)\varrho+\frac{\varrho_g}{\lambda^2_\varrho}.
	\end{split}
\end{align}
We assume that the derivative of the gravitational perturbation of the dilaton field is continuous on the plane, i.\,e., $\left.\partial\varrho_g/\partial z\right|_{z=0}=\bar\beta_\text{S}g/c^2$ for all times $t$, and vanishes at the origin, i.\,e., $\varrho_g(0)=0$.
Under these boundary conditions and to lowest order in $\bar\varrho_0$ and $g z/c^2$, without cross terms, we obtain the gravitational perturbation $\varrho_g(z)=\bar\beta_\text{S}g\lambda_\varrho\sin{\left(\left[1-\frac{g z}{2c^2}\right]z/\lambda_\varrho\right)}/c^2$.
Assuming a light dilaton $\lambda_\varrho\gg z$, implying that its Compton wave length is much larger than the extension of the experiment, the field 
\begin{align}
    \varrho(t,z)=\bar\varrho_0\cos{\left(\omega_\varrho t-k_\varrho z+\phi_\varrho\right)}+\bar\beta_\text{S}g z/c^2 
\end{align} 
solves the modified Einstein equations for regions outside the time-independent, homogeneous massive plane at $z=0$.
\bibliography{MaxwellBib}

\end{document}